\pdfoutput=1
\documentclass[twocolumn,
               showpacs,
               preprintnumbers,
               nofootinbib,
               prd,
               superscriptaddress,
               10pt,
               notitlepage,
               aps]{revtex4-2}

\usepackage{graphicx,amssymb,amsmath,amsthm,amsfonts,epsfig,mathtools}

\usepackage[linktocpage]{hyperref}
\usepackage[usenames,dvipsnames]{color}
\usepackage{epstopdf}
\usepackage{dsfont}
\usepackage{empheq}
\hypersetup{colorlinks=true, citecolor=NavyBlue,
            linkcolor=NavyBlue, urlcolor=NavyBlue}
\usepackage{mathrsfs}
\usepackage{tensor}
\usepackage{mathtools}
\usepackage{amsbsy}
\usepackage{bm}
\usepackage{tipa}
\usepackage[utf8]{inputenc}
\usepackage{microtype}

\usepackage{soul}

\newcommand{\nn}{\nonumber}

\newcommand{\be}{\begin{equation}}
\newcommand{\ee}{\end{equation}}
\newcommand{\ba}{\begin{align}}
\newcommand{\ea}{\end{align}}
\newcommand{\vp}{\varphi}

\newcommand{\spc}{\phantom{a}}

\newcommand{\meff}{\mu_{\text{eff}}}

\newcommand{\del}{\partial}
\newcommand{\dd}{{\rm d}}

\newcommand{\dx}{{\rm d}^4x \sqrt{-g}}
\newcommand{\jg}{\tilde{g}}

\newcommand{\dpsi}{\delta \psi}
\newcommand{\dA}{\delta A}
\newcommand{\argp}{(\theta,\phi)}
\newcommand{\argtr}{(t,r)}
\newcommand{\lm}{{\ell m}}

\newcommand{\ag}{\alpha}
\newcommand{\bg}{\beta}
\newcommand{\cg}{\gamma}

\newcommand{\tz}{\textctyogh}

\newcommand{\subvec}{{\rm v}}
\newcommand{\subdef}{{\mbox{\tiny DEF}}}
\newcommand{\mv}{\mu_{\rm v}}

\begin{document}
\title{The ghost of vector fields in compact stars}
\begin{abstract}
Spontaneous scalarization is a mechanism that allows a scalar field
to go undetected in weak gravity environments and yet develop a
nontrivial configuration in strongly gravitating systems.
At the perturbative level it manifests as a tachyonic instability
around spacetimes that solve Einstein's equations.
The endpoint of this instability is a nontrivial scalar field configuration
that can significantly modify a compact object's structure and can produce
observational signatures of the scalar field's presence.
Does such a mechanism exists for vector fields?
Here we revisit the model that constitutes the most straightforward
generalization of the original scalarization model to a vector field
and perform a perturbative analysis.
We show that a ghost appears as soon as the square of
the naive effective mass squared becomes negative anywhere.
This result poses a serious obstacle in generalizing spontaneous scalarization to vector fields.
\end{abstract}

\author{Hector O. Silva}
\email{hector.silva@aei.mpg.de}
\affiliation{Max Planck Institute for Gravitational Physics (Albert Einstein Institute),
Am M\"uhlenberg 1, 14476 Potsdam, Germany}

\author{Andrew Coates}
\email{acoates@ku.edu.tr}
\affiliation{Department of Physics, Ko\c{c} University,
Rumelifeneri Yolu, 34450 Sariyer, Istanbul, Turkey}

\author{Fethi M. Ramazano\u{g}lu}
\email{framazanoglu@ku.edu.tr}
\affiliation{Department of Physics, Ko\c{c} University,
Rumelifeneri Yolu, 34450 Sariyer, Istanbul, Turkey}

\author{Thomas P. Sotiriou}
\email{Thomas.Sotiriou@nottingham.ac.uk}
\affiliation{School of Mathematical Sciences \& School of Physics and Astronomy,
University of Nottingham, University Park, Nottingham, NG7 2RD, United Kingdom}

\maketitle

\section{Introduction}
\label{sec:intro}

The first gravitational wave (GW) signal from a compact binary
coalescence detected by the LIGO-Virgo collaboration~\cite{LIGOScientific:2016aoc}
in 2015 opened a new vista into the nonlinear and highly dynamical regime of gravity.
Moreover, and perhaps more excitingly, GWs now allow us to probe (or constrain)
new physics beyond GR and the Standard Model~\cite{Berti:2015itd,Yunes:2016jcc,Barack:2018yly,Nair:2019iur,LIGOScientific:2020tif,Perkins:2021mhb}.
This had so far been limited to
astronomical probes either in the weak gravitational field and slow velocity
in our Solar System or in the strong gravitational field, but small velocity and
large separation regime of binary pulsars~\cite{Will:2014kxa,Wex:2020ald}.

In this context, a particularly appealing new physics scenario is
one where new fundamental fields lie ``dormant'' in weak-gravity
environments and yet manage to have significant effects in strongly-gravitating
bodies and systems.
The prototypical theory that achieves this was first introduced by Damour and Esposito-Far\`ese~\cite{Damour:1993hw,Damour:1996ke}
and involves a massless scalar field $\varphi$.
The theory is described by the action
\begin{align}
S_\subdef &= \frac{1}{16 \pi} \int \dx\, \left(R - 2\, \nabla_{\mu}\varphi \,\nabla^{\mu}\varphi\right)
\nn \\
&\quad
+ S_{\rm m} \left[ \Psi_{\rm m};\, \Omega_{\subdef}^2(\varphi) \, g_{\mu\nu} \right],
\label{eq:action_def}
\end{align}
where $g$ is the metric determinant, $R$ is the Ricci scalar, $S_{\rm m}$ is the
action of matter fields $\Psi_{\rm m}$, which couple to $\Omega_{\subdef}^2(\vp) \, g_{\mu\nu}$, with
$\Omega_{\subdef} = \exp\left( \beta \varphi^2 / 2 \right)$; $\beta$
being a dimensionless constant.

The scalar field satisfies the field equation
\begin{equation}
    \Box \varphi
    = - 4 \pi \beta\, \Omega_{\subdef}^{4} \tilde{T}\, \varphi,
    \label{eq:box_def}
\end{equation}
where $\tilde{T}$ is the trace of the matter field's energy-momentum tensor.
Equation~\eqref{eq:box_def} clearly admits a vanishing scalar field as a solution.
However this is not the only solution for a given matter configuration.
Linearized scalar field perturbations $\delta \varphi$ on
the background of a neutron star can be shown to obey a wave equation
\begin{equation}
(\Box - \meff^2) \delta \varphi = 0,
\quad
\meff^2 = - 4 \pi \beta \, \Omega_{\subdef}^{4} \tilde{T},
\label{eq:meff_def}
\end{equation}
where $\meff^2$ is a position dependent \emph{effective mass squared}.
For a neutron star described by a perfect fluid, $\tilde{T} = 3 \tilde{p} - \tilde{\varepsilon}$
(where $\tilde{p}$ is the pressure and $\tilde{\varepsilon}$ the total energy density).
Typically $\tilde{T}<0$ and thus these perturbations can become tachyonic when $\beta < 0$~\cite{Harada:1997mr,Chiba:1997ms}, with only a weak dependence on the equation
of state~\cite{Novak:1998rk,Silva:2014fca,AltahaMotahar:2017ijw}.
Numerical simulations show that this linear instability is ultimately nonlinearly quenched and
thus the star becomes \emph{spontaneously scalarized}.
Due to Eq.~\eqref{eq:box_def}, these scalarized stars coexist with the
GR solutions [defined as stars with $\varphi = 0$]
and, importantly, are energetically favored: thus they can form
dynamically from stellar collapse~\cite{Novak:1998rk,Novak:1999jg,Gerosa:2016fri,Sperhake:2017itk,Rosca-Mead:2020ehn}
or in neutron star
binaries~\cite{Barausse:2012da,Palenzuela:2013hsa,Shibata:2013pra,Taniguchi:2014fqa,Sennett:2016rwa,Sennett:2017lcx}.

The Damour--Esposito-Far\`ese scalarization model cannot lead to black holes scalarization unless the latter is induced by surrounding matter~\cite{Cardoso:2013fwa,Cardoso:2013opa}. However, more general models which fashion couplings with the Gauss-Bonnet invariant, have been shown to lead to black hole scalarization, controlled by the mass~\cite{Doneva:2017bvd,Silva:2017uqg} or by the spin of the black hole~\cite{Dima:2020yac,Herdeiro:2020wei,Berti:2020kgk} and can take place in stellar collapse~\cite{Kuan:2021lol}.
Black hole scalarization can also have potentially observable effects in
binary black hole binaries~\cite{Silva:2020omi,East:2020hgw,East:2021bqk}
and be induced by other curvature scalars, such as the Pontryagin invariant~\cite{Gao:2018acg,Doneva:2021dcc}.
The instability leading to scalarization can also be understood from a
quantum field theory perspective, see e.g., Refs.~\cite{Lima:2010na,Mendes:2013ija,Pani:2010vc}.

A different type of generalization of the Damour--Esposito-Far\`ese mechanism that has
been explored is to extend it to vector fields~\cite{Ramazanoglu:2017xbl}.
Inspired by~\cite{Damour:1993hw,Damour:1996ke}, Ref.~\cite{Ramazanoglu:2017xbl} studied the action
\begin{align}
S_\subvec &= \frac{1}{16 \pi} \int \dx\, \left( R
- F_{\mu\nu} F^{\mu\nu}
- 2 \mv^2 A_{\mu}A^{\mu} \right)
\nonumber \\
&\quad+ S_{\rm m} \left[ \Psi_{\rm m};\, {\Omega}^2_{\subvec}(A_{\mu}) \, g_{\mu\nu}\right],
\label{eq:action}
\end{align}
where $F_{\ag\bg} = \nabla_{\ag} A_{\bg} - \nabla_{\bg} A_{\ag}$
is the antisymmetric Faraday tensor and $A_{\ag}$ is a vector
field with bare mass $\mv$.\footnote{Ref.~\cite{BeltranJimenez:2013fca} is an earlier study of a similar theory that is mainly concerned with cosmology.}
In analogy with $\Omega_{\subdef}$, the conformal factor is chosen as
$\Omega_{\subvec} = \exp\left( \beta A_{\mu} A^{\mu} / 2 \right)$,
where $\beta$ is a free parameter of the theory.
The field equation of $A_{\mu}$ is
\begin{equation}
    \nabla_{\mu} F^{\mu\alpha} = (\mv^2 - 4 \pi \beta \, \Omega_{\rm v}^{4} \tilde{T} ) A^{\alpha}.
    \label{eq:eom_vt}
\end{equation}
This equation promotes the bare mass $\mv$ of the Proca field
to what appears to be an effective mass squared
$\meff^2 = \hat{z} \mv^2$,
where
\begin{equation}
    \hat{z} = 1-4\pi ({\beta}/{\mv^2}) \Omega_{\rm v}^4 \tilde{T},
    \label{eq:def_zhat}
\end{equation}
for linearized vector field perturbations.
This effective mass squared can become negative in the presence of dense matter
as in the theory~\eqref{eq:action_def}.
This property is not specific to the theory~\eqref{eq:action}, and
is shared with other vector-tensor theories with curvature coupling
terms~\cite{Annulli:2019fzq,Ramazanoglu:2019gbz,Kase:2020yhw,Barton:2021wfj} or
disformal couplings~\cite{Ramazanoglu:2019jrr,Minamitsuji:2020pak}.

Based on the similarity between the field equations~\eqref{eq:box_def} and~\eqref{eq:eom_vt}
it is natural to expect that in the theory~\eqref{eq:action}, $A_{\mu}$ could also become
\emph{tachyonically unstable} around sufficiently compact neutron stars and
a \emph{spontaneous vectorization} mechanism exists.
Although nonlinear vectorized neutron star solutions have indeed been shown to exist in~\cite{Ramazanoglu:2017xbl}, the perturbative manifestation of vectorization has not been explored yet. This leaves a number of open questions unanswered.
In particular, a massive vector $A_{\mu}$ is known to propagate an additional longitudinal degree of freedom. What is its role in this process? Could vectorization be scalarization in disguise to some extent? More generally, can it be understood intuitively, as is the case for scalarization, as a tachyonic instability quenched by nonlinearities?
Answering these questions is important from  a model-building perspective, but also from a phenomenological perspective. They become even more pressing once one observes that, intuitively speaking, the aforementioned longitudinal mode gets contributions in its kinetic term from the $A^\mu A_\mu$ terms in the action. That kinetic term  will therefore have a nontrivial structure, which in turn raises doubts about whether this mode is well behaved.

Motivated by these questions, here we revisit the model of Ref.~\cite{Ramazanoglu:2017xbl} from a perturbative perspective and indeed uncover a \emph{ghost instability}.
Therefore vectorization appears to be fundamentally different from scalarization.
It also strongly suggests that the time-evolution problem of a star undergoing vectorization is potentially ill-posed, casting serious doubts on the viability of this theory and other related ones.
Combined with the work~\cite{Garcia-Saenz:2021uyv} which also found ghost (and gradient)
instabilities in generalized Proca theories in compact object backgrounds, our work
raises serious questions about the possibility to generalize the original mechanism of Damour~and~Esposito-Far\`ese beyond scalars since all proposed vectorization theories feature at least ghost instabilities.

The remainder of this paper explains how we arrived at these conclusions.
In Secs.~\ref{sec:theory} and~\ref{sec:field_eqs} we review the model
introduced in~\cite{Ramazanoglu:2017xbl}, restore gauge invariance by
performing the Stuckelberg trick and analyse the resulting field equations.
In Sec.~\ref{sec:pert} we linearize the theory's action in the background of a
nonrotating, spherically symmetric star and show how ghost instability appears.
In Sec.~\ref{sec:nonlinear} we lift the assumption of linearized gauge field
perturbations and consider the complete set of field equations. We show how
ghosts, which first went unnoticed in~\cite{Ramazanoglu:2017xbl}, arise.
In Sec.~\ref{sec:conclusions} we summarize our main results.

We work with geometrical units $c = G = 1$ and use the $(-,+,+,+)$ metric signature.
Symmetrization of indices is defined as $A_{(\ag\bg)} \equiv (A_{\ag\bg}+
A_{\bg\ag})/2$ and the antisymmetrization by $A_{[\ag\bg]} \equiv (A_{\ag\bg} -
A_{\bg\ag})/2$.

\section{A model for spontaneous vectorization with gauge symmetry}
\label{sec:theory}

Action~\eqref{eq:action} has been constructed in analogy with~\eqref{eq:action_def},
but a caveat of the resulting tensor-vector theory is absence of gauge invariance under
$A_{\ag} \to A_{\ag} + \del_{\ag} \lambda$
($\lambda$ being a scalar function) due to the mass term $\mv^2 A_{\mu}A^{\mu}$.
To restore gauge invariance, and at the same time more easily investigate the different degrees of freedom in the vector field, we can apply the Stueckelberg
trick~\cite{Ruegg:2003ps}. It consists of introducing a scalar field
$\psi$ (the Stueckelberg field) through the substitution
\begin{equation}
A_{\alpha} \to A_{\alpha} + \mv^{-1} \nabla_{\alpha} \psi,
\label{eq:trick}
\end{equation}
which results in a scalar-vector-tensor theory,
\begin{align}
S &= \frac{1}{16 \pi} \int \dx\, [ R - F_{\mu\nu} F^{\mu\nu}
\nonumber \\
&\left.\quad - \, 2 g^{\mu \nu} (\mv A_{\mu} + \nabla_{\mu} \psi)
(\mv A_{\nu} + \nabla_{\nu} \psi) \right]
\nonumber \\
&\quad+
S_{\rm m} \left[ \Psi_{\rm m};\, {\Omega}^2 \, (A_{\ag},\nabla_{\ag}\psi)\, g_{\mu\nu}\right],
\label{eq:action_stu}
\end{align}
with conformal factor
\begin{equation}
\ln {\Omega} = \frac{\beta}{2 \mv^2} g^{\mu \nu}
(\mv A_{\mu} + \nabla_{\mu} \psi)
(\mv A_{\nu} + \nabla_{\nu} \psi).
\label{eq:conf_stu}
\end{equation}
The theory is now  gauge invariant under the simultaneous transformations:
\begin{equation}\label{eq:gaugetrans}
A_{\ag} \to A_{\ag} + \nabla_{\ag} \lambda, \qquad
\psi \to \psi - \mv \lambda.
\end{equation}
We see that $\psi$ can be set to zero by a suitable choice of $\lambda$ and thus the action~\eqref{eq:action} is a gauge-fixed version of action~\eqref{eq:action_stu}.

Indeed, for $\beta=0$ the conformal factor $\Omega$ becomes unity
and we recover the Stueckelberg theory minimally coupled to gravity (see e.g.,~\cite{Belokogne:2015etf}).
If we fix a gauge where $\psi = 0$ (we
call this the ``Proca gauge''), we obtain the nonminimally coupled Einstein-Proca theory of Ref.~\cite{Ramazanoglu:2017xbl}.
If we instead take $\mv \to 0$ we obtain the Einstein-Maxwell theory with the
addition of a scalar field.
In Proca theory the $\mv \to 0$ limit has
an apparent discontinuity of the longitudinal polarization mode of $A_{\ag}$.
In the ``Stueckelberged'' version of the same theory, the $\mv \to 0$ limit is manifestly continuous and corresponds
to the decoupling between $\psi$ and $A_{\ag}$ (the latter associated
with the usual Maxwell theory).
Note that, when $\beta\neq 0$, maintaining regularity of $\Omega$ requires that $\beta$ approaches zero
at least as fast as $\mv^2$ when taking the limit $\mv \to 0$.

The action~\eqref{eq:action_stu}
is written in the Einstein frame (thus we call
$g_{\ag\bg}$ the Einstein frame metric). We will refer to
$\jg_{\ag\bg} = \Omega^2 g_{\ag\bg}$ as the Jordan frame metric.
We will use tildes to denote objects in the Jordan frame, some of
which, as $\tilde{T}$, already appeared in the Introduction.

\section{The field equations}
\label{sec:field_eqs}

The field equations of the theory can be obtained by varying the action~\eqref{eq:action_stu}
with respect to $\psi$, $A_{\ag}$ and $g^{\ag\bg}$:
\begin{align}
\Box\psi &= - \mv \nabla_{\mu} A^{\mu}
+ (4 \pi / \mv) \, \nabla_{\mu} (\alpha^{\mu}_{\rm A} \Omega^4 \tilde{T}),
\label{eq:box_sca_e}
\\
\nabla_{\mu} F^{\mu\ag} &=
\mv g^{\mu\ag}(\mv A_{\mu} + \del_{\mu} \psi)
- 4\pi\alpha_{\rm A}^{\ag}\, {\Omega}^4\, \tilde{T},
\label{eq:divfmunu}
\\
G_{\ag\bg} &= 8 \pi \left(T^{\rm e}_{\ag\bg}
+ T^{\rm s}_{\ag\bg}
+ T_{\ag\bg}\right),
\label{eq:einstein_eq}
\end{align}
where,
\begin{equation}
\alpha^{\mu}_{\rm A} \equiv \frac{\del\ln \Omega}{\del A_{\mu}}
= \mv \frac{\del\ln \Omega}{\del (\nabla_{\mu} \psi)}
= \frac{\beta}{\mv}(\mv A^{\mu} + \nabla^{\mu}\psi),
\end{equation}
and we defined the individual energy-momentum contributions from
``pure electromagnetic'' theory $T^{\rm e}_{\mu\nu}$ and from the
``Stueckelberg contribution'' to the action $T^{\rm s}_{\mu\nu}$,
\begin{align}
T^{\rm e}_{\ag\bg} &= \frac{1}{4\pi}\left(F_{\mu \ag} F_{\nu \bg} g^{\mu\nu}
- \tfrac{1}{4} g_{\ag\bg} F^{\mu\nu} F_{\mu\nu}\right),
\label{eq:tmunu_em}
\\
T^{\rm s}_{\ag\bg} &= \frac{1}{4\pi}\left[\left( \mv A_{\ag} + \nabla_{\ag}\psi\right)
\left( \mv A_{\bg} + \nabla_{\bg}\psi\right)\right.
\nonumber \\
&\left.\quad-\tfrac{1}{2} g_{\ag\bg}
\left( \mv A^{\mu} + \nabla^{\mu}\psi\right)
\left( \mv A_{\mu} + \nabla_{\mu}\psi\right)\right].
\label{eq:tmunu_stu}
\end{align}
The Jordan frame energy-momentum tensor of matter fields and its
trace are defined as
\begin{equation}
{\tilde T}_{\ag\bg}
\equiv -\frac{2}{\sqrt{-g}}\frac{\delta S_{\rm m}}{\delta \jg^{\ag\bg}},
\quad \textrm{and} \quad
 {\tilde T} \equiv \jg^{\mu\nu}{\tilde T}_{\mu\nu}.
\end{equation}
We also have by construction:
\begin{equation}
\nabla_{\ag}F_{\bg\cg} +
\nabla_{\cg}F_{\ag\bg} +
\nabla_{\bg}F_{\cg\ag} = 0.
\end{equation}

Going back to Eq.~\eqref{eq:divfmunu} and due to
$\nabla_{\mu}\nabla_{\nu} F^{\mu\nu} = 0$, it is convenient to define a current $j^{\ag}$
as:
\begin{equation}
j^{\ag} = \mv g^{\mu\ag}(\mv A_{\mu} + \nabla_{\mu} \psi)
- 4\pi\alpha_{\rm A}^{\ag}\, \Omega^4\, \tilde{T},
\label{eq:def_j}
\end{equation}
which is conserved
\begin{equation}
\nabla_{\mu} j^{\mu} = 0.
\label{eq:cons_j}
\end{equation}
In terms of $j^{\ag}$ we have:
\begin{equation}
\mv^2 \nabla_{\mu} A^{\mu} =
-\nabla_{\mu}
(
\mv\nabla^{\mu}\psi - 4\pi\alpha_{\rm A}^{\mu}\,\Omega^4\, {\tilde T}
).
\label{eq:current_explicit}
\end{equation}
In the absence of matter (${\tilde T} = 0$) and in the Proca gauge ($\psi = 0$),
Eq.~\eqref{eq:current_explicit} becomes the Lorenz constraint on $A_{\ag}$ of Proca theory.
Thus, the field equation for $\psi$ [cf.~Eq.~\eqref{eq:box_sca_e}]
and the Lorenz constraint on $A_{\mu}$ are tightly connected.

We can see the first sign of the ghost by introducing a third metric,
\begin{equation}
\overline{g}_{\alpha\beta}= \hat{z}^{-1} \, g_{\alpha\beta},
\end{equation}
in terms of which the scalar field equation becomes
%
\begin{equation} \label{eq:box_psi_eff_metric}
\overline{\Box}\psi = -\overline{g}^{\mu\nu}
[ \mv \overline{\nabla}_\mu A_\nu +\tfrac{1}{2}
( \overline{\nabla}_\mu \log \hat{z} )
( \overline{\nabla}_\nu \psi+\mv A_\nu )
].
\end{equation}
This third metric can, in principle, have a signature change in some parts of the spacetime due to the $\hat{z}^{-1}$ term. If this happens, the field will be a ghost in at least some region compared to any field which is coupled to a fixed signature metric.
Another potential problem is the fact that this metric changes sign by diverging, rather than crossing zero, in a similar vein
discussed in~\cite{Minamitsuji:2016hkk,Ventagli:2020rnx}.
It is unclear whether there is a rectification for such a problem, or, worse,
whether the theory can evolve from a state where this metric has a fixed signature to another where the signature changes.

It is also instructive to consider the limit \(\mv= 0\), with \(\beta \to 0\) as fast as \(\mv^2\).  In this limit, the Stueckelberg field $\psi$ is no longer affected by gauge transformations, so $A_\mu$ becomes a gauge field. Then $A_\mu$ smoothly decouples from $\psi$ and the matter fields. However, there is still coupling to gravity and $\psi$ continues to be coupled to matter.
In particular, Eq.~\eqref{eq:box_psi_eff_metric} becomes,
\begin{equation}
    \overline{\Box} \psi = - \tfrac{1}{2} \, \overline{g}^{\mu\nu}
    (\overline{\nabla}_\mu \log \hat{z})
    (\overline{\nabla}_\nu \psi),
\end{equation}
note that $\hat{z}$ does not depend on $A_{\alpha}$ here.
%
So, $\psi$ will become a ghost when the $\bar{g}_{\mu\nu}$ metric changes signature and, as it is coupled to gravity and matter, its ghostly nature is physical.
This same procedure is used in the Stueckelberg picture of Proca theory to show that $\psi$ and $A_\mu$
decouple and hence there is no discontinuity as \(\mv= 0\) (i.e., no degree of freedom disappears).
In this setting, one has $\beta=0$, flat spacetime, and no matter.

One may object that \(\psi\) can be completely removed by a gauge choice
such as the Proca gauge $\psi=0$, and thus the ghost can be exorcised.
For this reason we will use the rest of the paper to assuage any doubts.
We will begin by examining the quadratic Lagrangian
for scalar-vector perturbations around a neutron star GR solution.
Doing so we will find there exists a gauge invariant scalar field that suffers the same problems.

\section{Perturbative analysis}
\label{sec:pert}

\subsection{Background spacetime and overview of the calculation}
\label{sec:pert_outline}

In this section we explore the test-field limit of our theory, where
we study the dynamics of $\psi$ and $A_{\ag}$ in a background corresponding to
a stellar solution of Einstein's field equations, i.e., a solution
of the Tolman--Oppenheimer--Volkoff (TOV) equations~\cite{Tolman:1939jz,Oppenheimer:1939ne}
whose line element we write as
\begin{equation}
\dd s^2 = - e^{\nu} \dd t^2 + \frac{r}{r-2\mu}\dd r^2 +
r^2 (\dd \theta^2 + \sin^2\theta\,\dd\phi ),
\label{eq:line_element}
\end{equation}
where $\nu$ (lapse) and $\mu$ (mass function) are functions of
the radial coordinate $r$ only.

In Sec.~\ref{sec:harmonics} we will linearize the field equations for small field perturbation~$\delta \psi$ and $\delta A_{\mu}$ at the level of the field equations~\eqref{eq:box_sca_e}--\eqref{eq:divfmunu}, and show how the ghost arises in this background.
Then, in Sec.~\ref{sec: quad_lagrangian}, we reach the same conclusion
by directly perturbing the Lagrangian, by expanding it
to second order in the fields on the same background.

\subsection{Linearized field equations}
\label{sec:harmonics}

We are interested in
studying the dynamics of $\dA_{\ag}$ and $\dpsi$
propagating on the background line element~\eqref{eq:line_element}.
To proceed we decompose $\dpsi$ and $\dA_{\ag}$ in scalar and vector harmonics
respectively. This is the convenient basis to expand scalar and vector fields
on the unit two-sphere and, thus, in problems with spherical symmetry.
We follow closely the presentation by Rosa and Dolan~\cite{Rosa:2011my},
but with a slightly different normalization.
More specifically, we write  $\dpsi$ as
\begin{equation}
\dpsi = \frac{1}{r}\sum_{\lm}\sigma_\lm\argtr Y_\lm\argp,
\end{equation}
where $Y_{\ell m} = Y_\lm\argp$ are the spherical harmonics
with $\ell = 0, 1, 2 \dots$, and $|m| \leqslant \ell$.
For the vector perturbations, we decompose $\dA_{\ag}$ as
\begin{equation}
\dA_{\ag} = \frac{1}{r}\sum^{4}_{i=1}\sum_\lm
c_{i} u^\lm_{(i)}(t,r) Z^{(i)\lm}_{\ag}\argp,
\end{equation}
where $c_1 = c_2 = 1$, $c_3 = c_4 = 1/\sqrt{\ell(\ell + 1)}$, and $Z^{(i)\lm}_{\ag}$ are
the vector harmonics given by
\begin{align}
Z^{(1)\lm}_{\ag} &= [1,0,0,0]Y_\lm,
\\
Z^{(2)\lm}_{\ag} &= [0, 1,0,0]Y_\lm,
\\
Z^{(3)\lm}_{\ag} &= \frac{r}{\sqrt{\ell(\ell+1)}}
[0,0,\del_{\theta},\del_{\phi}]Y_\lm,
\\
Z^{(4)\lm}_{\ag} &= \frac{r}{\sqrt{\ell(\ell+1)}}
[0,0,\csc\theta \del_{\phi}, -\sin\theta\del_\theta]Y_\lm.
\end{align}
These functions are orthonormal
when integrated on the unit two-sphere,
according to the inner product,
\begin{equation}
\int
(Z^{(i)\lm}_{\mu})^{\ast} \eta^{\mu\nu}
Z_{\nu}^{(i')\ell'm'}
\sin\theta\,\dd\theta\,\dd\phi
= \delta_{ii'}\delta_{\ell\ell'}\delta_{mm'},
\end{equation}
where $\eta^{\ag\bg} \equiv {\rm diag}[1,\, 1,\, (1/r^2),\, 1/(r^2\sin^2\theta)]$.

Under parity
inversion ${\bm x} \to -{\bm x}'$ (or equivalently, in spherical coordinates,
$\theta \to \pi - \theta$ and $\phi \to \phi + \pi$), the first three
harmonics ($i=1, 2, 3$) pick a factor of $(-1)^\ell$, while the
fourth ($i=4$) picks a factor of $(-1)^{\ell+1}$. We follow the
literature convention and call the former ``even parity''
modes and the latter ``odd parity''
modes.
The scalar perturbation $\dpsi$ is of even parity.

At this point it will be useful to follow a similar procedure to~\cite{Garcia-Saenz:2021uyv}.
We expand the Stueckelberged action~\eqref{eq:action_stu} around a GR
solution to second order in the test field approximation and find,
\begin{align}
S_{2} \left[\delta A, \delta \psi\right]
&=\frac{1}{4 \pi}\int\dx \,
[ 2 (\nabla^\nu\delta A^\mu) (\nabla_{[\mu} \delta A_{\nu]}) \nonumber\\
& - z\left(\mv \delta A_\mu + \nabla_\mu \delta\psi\right) \left(\mv \delta A^\mu +\nabla^\mu \delta\psi\right) ],
\nn \\
\label{eq:s2_full}
\end{align}
where,
\begin{equation}
z = 1 - 4 \pi (\beta / \mv^2)  \tilde{T},
\label{eq:z}
\end{equation}
which is unity outside the star, where $\tilde{T} = 0$.
Note that we could have arrived at an action
of this form by using the Stueckelberg trick in the Proca Lagrangian with
a ``dressed mass'' $z \mv^2$.
Therefore, the results of this section apply to any theory whose
quadratic Lagrangian can be put in this form, i.e., where one would naively
expect just a screened Proca field prone to develop a tachyonic instability.
Substituting the decompositions of $\delta A_{\alpha}$ and $\psi$ in harmonics,
results in a Lagrangian, with even and odd-parity sector decoupled from one another.
We look at each sector next.

\subsection{Monopolar even-parity quadratic Lagrangian}
\label{sec: quad_lagrangian}

We first focus on the monopole perturbations ($\ell = m = 0$),
which have the lowest instability threshold and belong to the even-parity sector.
Since $Y_{00} = \text{constant}$, only the $i=1,2$ vector harmonics
are defined~\cite{Rosa:2011my}.
This means that we would need to work with three
variables $\sigma_{00}$, $u^{(1)}_{00}$ and $u^{(2)}_{00}$,
\begin{subequations} \label{eq:l0_perts}
\begin{align}
\delta A_\alpha & =\frac{1}{2 \sqrt{\pi}} [ u_1 (t,r),\, u_2 (t,r),\, 0, \,0],\\
\delta\psi &= \frac{1}{2 \sqrt{\pi} r} \sigma (t,r),
\end{align}
\end{subequations}
where, to shorten the notation, we use $\sigma = \sigma_{00}$, $u_{1} = u^{(1)}_{00}$,
and $u_{2} = u^{(2)}_{00}$ hereafter.

Inserting Eqs.~\eqref{eq:l0_perts} in the action~\eqref{eq:s2_full}
and integrating over the angular coordinates leaves us with,
\begin{align}
S_2^{({\rm e})} &= \int\mathrm{d}t\mathrm{d}r\frac{e^{-\frac{\nu}{2}}\sqrt{r-2\mu}}{4\pi r^{5/2}}
\left\{\frac{z}{2 \mv^2} \left[\frac{r^3  (\mv r u_1 +  \dot{\sigma})^2}{r-2\mu}
\right.
\right.
\nn\\
&\quad \left. \left. -\,e^{\nu}( \sigma - r(\mv r u_1 + \sigma'))^2 \vphantom{ \frac{r^3}{r-2\mu}}\right]+
\frac{r^4}{2} [u_1'-\dot{u}_2]^2
\right\},
\nn \\
\label{eq:l0Action}
\end{align}
where we defined \((\cdot)' = \partial_r (\cdot) \) and \(\dot{(\cdot)} = \partial_t (\cdot)\). It can be readily verified that under the gauge transformation~\eqref{eq:gaugetrans} with \(\lambda = l/(2 \sqrt{\pi} r)\) that,
\begin{align}
\sigma \to \sigma - \mv l,
\,\,\,\,
\{u_1, u_2\} \to \{ u_1, u_2 \} + \{ \dot{l}/r, (l/r)' \},
\end{align}
and that the action~\eqref{eq:l0Action} is invariant under this transformation.
In fact, it can be
verified that, the combination,
\begin{equation}
\Phi=\dot{u}_2-u_1',
\end{equation}
is itself gauge invariant (proportional to the $\ell=0$ component of the electric field).
If we introduce the auxiliary field \(\phi\) such that, on shell, \(\phi=r^2\Phi\), we can rewrite Eq.~\eqref{eq:l0Action} as,
\begin{align}
S_2^{({\rm e})} &= \int\mathrm{d}t\mathrm{d}r\frac{e^{-\frac{\nu}{2}}\sqrt{r-2\mu}}{4\pi r^{5/2}}
\left\{ \frac{z}{2 \mv^2} \phantom{\frac{1}{1}}
\right.
\nn\\
&\quad
\left.
\times \left[- e^{\nu} (\sigma - r(\mv r u_2+ \sigma'))^2 \vphantom{ \frac{r^3}{r-2\mu}}
+
\frac{r^3 (\mv r u_1 + \dot{\sigma})^2}{r-2\mu}\right]
\right.
\nn \\
&\quad + \left. \frac{1}{2} \phi\left(2 r^2 \Phi-\phi\right)
\right\}.
\label{eq:l0auxAction}
\end{align}
In this formulation, $\phi$, $u_1$ and $u_2$ are all nondynamical: their equations of motion can be solved algebraically in the form, e.g.,
\begin{equation}
    u_1=u_1[u_2,\partial u_2,\phi, \partial \phi,\sigma, \partial \sigma].
\end{equation}
We can then replace this solution directly into the action, ``integrating out" whichever field.
Integrating out \(u_1\) and \(u_2\) one arrives at an action that is a functional of \(\phi\) alone (all terms involving $\sigma$ cancel). This transfers all of the dynamics from $\sigma$ to $\phi$.
The resulting action has the form,
\begin{align}
S_2^{({\rm e})} &= \int\mathrm{d}t\mathrm{d}r
\frac{e^{-\frac{\nu}{2}}\sqrt{r-2\mu}}{4\pi r^{5/2}}
\left\{
\frac{1}{2 z}
\left[
\vphantom{\frac{1}{2}}
e^{-\nu} \dot{\phi}^2- \left(1 - \frac{2\mu}{r} \right) \phi'^2\right.\right.
\nn\\
&\quad
\left.
\left.
-\frac{2 C_\times}{z r^2}\phi\phi' +\left(-z + \frac{z' C_1}{z r^2}+\frac{C_2}{4 r^3}\right) \phi^2\right]\right\},
\label{eq:quad_lagrangian_phi}
\end{align}
where
\begin{align}
C_\times &= r(r-2\mu) z'
\nn \\
&\quad + z [ (r-2\mu)(4+ r \nu')- 2 r(1- 2\mu') ],
\\
C_1 &= \nu' r(r-2\mu) + 2 r \mu' - 2 \mu,
\\
C_2 &= \frac{ r^2(1-2\mu')^2}{r-2\mu} - \{6 r(1- 2 \mu')(3 + r \nu')+ 8 r^2 \mu''
\nonumber\\
&\quad  - \, (r-2\mu)\left[17+r\nu'(14+r\nu')-4 r^2\nu''\right] \}.
\end{align}

We see immediately that the sign of the kinetic contribution changes if \(z\) does (and also diverges when \(z\) crosses \(0\)).
That is, we have shown that, in this situation, there is a gauge invariant statement
of the problems discussed in Sec.~\ref{sec:theory}, arising from Eq.~\eqref{eq:box_psi_eff_metric}.

\subsection{Odd-parity quadratic Lagrangian}
\label{sec:quad_lagrangian_odd}

Having identified the presence of a ghost in the even-parity sector, it is
natural to ask whether such ghosts also arise in the odd-parity sector, which contains
a single degree of freedom $u_4$, with multipole $\ell \geqslant 1$.
We find, after integration over the angular coordinates,
\begin{align}
S_{2}^{({\rm o})} &= \sum_{\ell = 1}^{\infty} \int \dd t \dd r
\frac{ e^{\frac{\nu}{2}}(1-2\mu/r)^{-\frac{1}{2}}}{4 \pi \ell(\ell + 1) }
\left\{
\vphantom{\frac{1}{2}}
e^{-\nu} (\dot{u}_{4})^{2}
\right.
\nn \\
&\quad \left.
- \left( 1 - \frac{2\mu}{r} \right) (u'_{4})^{2}
- \left[ \frac{\ell(\ell + 1)}{r^2} + z \mv^2 \right] u_{4}^2
\right\},
\nonumber \\
\end{align}
where we defined $u_{4} = u^{(4)}_{\ell 0}$ and set $m=0$ due to the background's
spherical symmetry.

Hence, we see that $u_{4}$ is prone to a tachyonic instability controlled by the
same effective mass squared $z \mv^2$ also responsible for inducing a ghost instability in the even-parity sector.
Indeed, the term between square brackets is the effective potential for
massive vector axial perturbations found in~\cite{Rosa:2011my}, Eq.~(13),
for $z = 1$.
We then conclude that the  axial sector can become tachyonic unstable, \emph{but} the
dominant effect occurs at lower multipole: the ghost instability in the even-sector.


\section{Unveiling the ghost in the Proca gauge}
\label{sec:nonlinear}

We have identified a ghost instability in the scalar sector of our theory,
however no ghosts were reported in the spontaneous vectorization theory
introduced in Ref.~\cite{Ramazanoglu:2017xbl}, or related theories investigated
in Refs.~\cite{Annulli:2019fzq,Ramazanoglu:2019gbz,Kase:2020yhw,Barton:2021wfj}.
In this section and related appendices, we will demonstrate that these theories contain divergent terms in their field equations \emph{irrespective of whether one uses the Stueckelberg trick to restore gauge symmetry or not}.

Recall that the Proca gauge ($\psi=0$) is equivalent to the spontaneous
vectorization theory of Ref.~\cite{Ramazanoglu:2017xbl}. Effectively,
this gauge undoes the Stueckelberg trick~\eqref{eq:trick} and we only need to consider
Eq.~\eqref{eq:eom_vt}.
Since there is no separate equation for $\psi$ in this picture and there are no
divergent terms in this field equation, it is unclear where the ghost
lurks. This is elucidated by considering the \emph{constraint equation}.

Since $\nabla_\mu \nabla_\nu F^{\mu\nu}=0$ still holds due to the antisymmetry of $F^{\mu\nu}$ in Eq.~\eqref{eq:eom_vt}, we obtain
\begin{align}
\nabla_\mu [ (\mv^2 - 4\pi \Omega^4 \beta \tilde{T}) A^\mu ]
 = 0.
\label{eq:X_constraint}
\end{align}
This is the generalized version of the $\nabla_\mu A^\mu = 0$ constraint for a
minimally coupled Proca field.

The puzzling aspect of Eq.~\eqref{eq:eom_vt} is that it does not have any explicit indication of a ghost,
however we now know from our discussion in Sec.~\ref{sec:field_eqs} that the constraint~\eqref{eq:X_constraint} given in the form of a conserved current
in Eq.~\eqref{eq:cons_j} is also crucial to understand the time evolution.
Indeed, the constraint imposes a time evolution for $A^0$ that will reveal the
ghost.\footnote{Note that $A^0$ is not a dynamical degree of freedom in the
standard Hamiltonian sense~\cite{Heisenberg:2018vsk}.
The zeroth-component of the equation of motion~\eqref{eq:eom_vt}, is not a
time-evolution equation; it imposes an elliptic constraint on $A^0$ in terms of
the other components of the vector and matter fields. However, one can
indirectly calculate how $A^0$ evolves in time through the evolution
of these other degrees of freedom, which can be obtained by the
constraint.}

Let us rewrite the constraint in terms of $\hat{z}$ [cf. Eq.~\eqref{eq:def_zhat}],
\begin{align}
\nabla_\mu \left( \hat{z} A^\mu \right) = 0.
\label{eq:constraint2}
\end{align}
We can convert the covariant derivatives to
partial derivatives to obtain
\begin{align}
\partial_0 ( \sqrt{-g} \hat{z} A^0 ) = -\partial_i ( \sqrt{-g} \hat{z} A^i ),
\label{eq:X_constraint3}
\end{align}
where $i$ runs over the spatial coordinates. We see that this time-evolution equation has divergent terms due to the behavior of $\hat{z}$, even if all fields other than $A^0$ are regular.
Outside any matter distribution $\hat{z}=1$ and we require $\hat{z}<0$ in some
part of spacetime if we want an astrophysical object to vectorize.
Since $\hat{z}$ is continuous, it has to vanish at some point. There is no
symmetry to ensure that $\sqrt{-g} \hat{z} A^0$ vanishes where $\hat{z}$ vanishes since $\hat{z}$ and its derivatives do
not vanish at the same spacetime points in general. This means, $A^0$ will generically diverge
even if $\sqrt{-g} \hat{z} A^0$ stays regular.
Alternatively, we can move the $\hat{z}$ term outside the
derivative on the left-hand side, which means that now the coefficient
of the leading time derivative of $A^0$ vanishes at certain points.
This means that the divergent terms we observed in the ghost instabilities of
$\psi$ manifest themselves \emph{not directly in the field equation~\eqref{eq:eom_vt},
but in the constraint equation~\eqref{eq:X_constraint}, or equivalently, in Eq.~\eqref{eq:X_constraint3}}.

The dynamics of $A^0$ implied by Eq.~\eqref{eq:X_constraint3} is first
order in time, thus not strictly of the same nature of the wave equation
obeyed by $\psi$. Nonetheless, the change of sign in the time derivative
leads to an analogous pathology.
This can be understood by recasting the
field equation~\eqref{eq:eom_vt} into an explicitly hyperbolic form.

Let us start by rearranging the constraint~\eqref{eq:constraint2} as
\begin{align}
\nabla_\mu \left( \hat{z} A^\mu \right) = 0\
\Rightarrow \ \nabla_\mu A^\mu = - A^\mu \, \nabla_\mu \ln|\hat{z}|
\label{eq:X_constraint2}
\end{align}
Next, we manipulate Eq.~\eqref{eq:eom_vt} as follows
\begin{align}
\hat{z} \mv^2 A_\alpha &= \nabla_\mu F^{\mu}_{\phantom{z}\alpha}, \nn \\
& =\nabla_\mu \nabla^\mu A_\alpha - \nabla_\mu \nabla_\alpha A^\mu,  \nn \\
&= \Box A_\alpha - \nabla_\alpha \nabla_\mu A^\mu
-R^{\mu}_{\phantom{z}\nu\mu\alpha} A^\nu, \nn \\
&=
\Box A_\alpha
+ \nabla_\alpha \left( A^\mu \nabla_\mu \ln|\hat{z}| \right)
-R_{\alpha\mu} A^\mu ,
\label{eq:vector_eom_mani}
\end{align}
where we related the commutator of two covariant derivatives to the Riemann
tensor in the third line, and used the constraint equation~\eqref{eq:X_constraint2}
in the fourth line.
We finally obtain
\begin{align}
\Box A_\alpha
+ ( \nabla_\mu \ln |\hat{z}| ) \nabla_\alpha A^\mu
= \mathcal{M}_{\alpha\mu} A^\mu,
\label{eq:vec_component}
\end{align}
where we defined the mass-squared tensor
\begin{align}
\mathcal{M}_{\alpha\beta} = \hat{z} \mv^2 g_{\alpha\beta}  + R_{\alpha\beta}
- \nabla_\alpha \nabla_\beta \ln |\hat{z}|.
\label{eq:mass-tensor}
\end{align}
We should be cautious about the fact that $\hat{z}$ contains $A^\alpha$ terms
[inside the conformal factor; cf.~Eq.~\eqref{eq:def_zhat}],
which, strictly speaking, means that $\nabla_\alpha \nabla_\beta \ln|\hat{z}| $ also belongs to the
principal part of the differential equation.
However, for
perturbative values of $A^\alpha$, such as in a fixed background calculation of Sec.~\ref{sec:pert},
this dependence can be ignored to leading order and $\mathcal{M}_{\alpha\beta}$ becomes a
proper mass-square tensor. Hence, Eq.~\eqref{eq:vec_component} can be viewed
as a generalized massive wave equation.

Equation~\eqref{eq:vec_component} has a divergent mass term due to
various factors of $\hat{z}^{-1}$ on its right-hand side. We have the
option of moving these factors to the left-hand side, which means
the principal part becomes $\hat{z} \Box A_\mu$. This is a
field equation prone to a ghost instability since $\hat{z}$ changes sign as we discussed before in Eq.~\eqref{eq:box_psi_eff_metric}. One can also
analyze the equation of motion for each vector harmonic,
which likewise leads to divergent effective mass terms.

The behavior of $\hat{z}$ is slightly modified for a vector field with
no intrinsic mass, $\mv=0$.
In this case Eq.~\eqref{eq:def_zhat}, and correspondingly Eq.~\eqref{eq:constraint2}, are modified as
\begin{align}
\hat{z}= - 4 \pi \beta \, \Omega_{\rm v}^{4} \tilde{T}
=- 4 \pi \beta \, \Omega_{\rm v}^{4} (3\tilde{p}-\tilde{\varepsilon}),
\label{eq:def_zhat2}
\end{align}
where we assume the neutron star matter to behave as a perfect fluid
with Jordan frame total energy density $\tilde{\varepsilon}$ and
pressure $\tilde{p}$ as before.
We see that $\hat{z}$ vanishes outside the star and
is generally negative within it; thus it
never crosses zero.
However, there are still divergences.

The first case of the divergence in the field equations for $\mv=0$ occurs at the surface of the neutron star.
The relevant part of the TOV equations for a spherically symmetric star is~\cite{Tolman:1939jz,Oppenheimer:1939ne},
\begin{align}
\frac{\dd \tilde{p}}{\dd r} =
- \frac{\tilde{\varepsilon} \mu}{r^2}
\left( 1 + \frac{\tilde{p}}{\tilde{\varepsilon}} \right)
\left( 1 + \frac{4\pi\tilde{p}r^3}{\mu} \right)
\left( 1 - \frac{2\mu}{r}\right)^{-1}.
\label{eq:tov_p}
\end{align}
In the outer layers of the star one has $\tilde{p} \ll \tilde{\varepsilon}$ and $4 \pi \tilde{p} r^3 \ll \mu$~\cite{Haensel:2007yy,Eksi:2015red}, which allows us to approximate Eq.~\eqref{eq:tov_p} as
\begin{equation}
\frac{\dd \tilde{p}}{\dd \varrho} = - \tilde{\rho} \, {\rm g},
\label{eq:tov_p_outer}
\end{equation}
where we approximated the total energy-density as equal
to the rest mass density ($\tilde{\varepsilon} \approx \tilde{\rho}$), introduced the proper radial length $\varrho$ [related to the coordinate radius $r$ as $\dd \varrho / \dd r = (1 - 2\mu / r)^{-1/2}$], and defined the ``local gravitational acceleration''
${\rm g} = (\mu / r^2) (1 - 2 \mu / r)^{-1/2}$~\cite{Haensel:2007yy}.

Focusing on the outer envelope of the star~\cite{Urpin:1979ApU}, we can approximate the spacetime as being Schwarzschild, i.e., $\mu \approx M$ and $\nu = \ln(1 - 2M/r_s)$
in Eq.~\eqref{eq:line_element}, where $M$ is the mass and $r_s$ the radius of the star.
We can further introduce a local proper depth $\textrm{\tz} = (R - r) ( 1 - 2 M / r_s)^{-1/2}$, in terms of which we can recast Eq.~\eqref{eq:tov_p_outer} as,
\begin{equation}
\frac{\dd \tilde{p}}{\dd \textrm{\tz}} = {\rm g}_{s} \, \tilde{\rho},
\label{eq:atmos}
\end{equation}
i.e., the equation of a plane-parallel atmosphere with a relativistic-corrected
surface gravity ${\rm g}_{s} = {\rm g}(r_s) = (M / r_s) (1 - 2M / r_s)^{-1/2}$.
In the outermost stellar layers, the main contribution to the pressure is
due to a nonrelativistic degenerate electron gas, for which Eq.~\eqref{eq:atmos}
can be solved exactly (see Ref.~\cite{Haensel:2007yy}, Sec.~6.9), yielding the scaling $\tilde{\rho} \propto \textrm{\tz}^{3/2}$ and, within the same assumptions, $\tilde{T} \propto - 4 \pi \beta \, \textrm{\tz}^{3/2}$.
This, in turn, means that $\nabla_\mu \ln |\hat{z}|$ and the effective mass diverge
on the surface, completing our argument.
The same reasoning can in principle be applied to other systems which have an interface of vacuum and matter, suggesting that any such interfaces would lead to a divergence in the vector field equations in general.
These divergences at the surface of the star are not exclusive to the
vector-tensor model considered here, but are known to also arise, albeit
with a different origin, in Palatini $f(R)$~\cite{Barausse:2007pn,Barausse:2007ys,Barausse:2008nm} and in Eddington-inspired Born-Infeld~\cite{Pani:2012qd} theories.
See also~\cite{Sotiriou:2008dh,Pani:2013qfa}.

The second case of divergence in the field equations for $\mv=0$ is
related to massive neutron stars.
Although $\tilde{T}$ is negative in general, it can switch sign and become positive
in the core of such stars for some equations of state (see e.g.,~\cite{Mendes:2014ufa,Palenzuela:2015ima,Mendes:2016fby,Podkowka:2018gib}).
This means that $\hat{z}$ vanishes somewhere inside the star [cf.~Eq.~\eqref{eq:def_zhat2}], where our previous results for the $\mv \neq 0$ case directly apply.

Overall, the above discussion provides a heuristic tool
to identify ghosts in spontaneous vectorization theories. If the
spacetime dependent $\meff^2$
vanishes in nonvacuum regions in a theory with field equation $\nabla_\mu F^{\mu\alpha} = \meff^2 A^\alpha$,
this generically leads to divergent terms in the
explicitly hyperbolic field equations. In other words, despite
the appearances and the naming we used, $\meff$ is \emph{not the
effective mass of all physical degrees of freedom}.
A careful analysis reveals that the true effective mass diverges as in Eq.~\eqref{eq:vec_component}, which was overlooked in the original
spontaneous vectorization theory of Ref.~\cite{Ramazanoglu:2017xbl}
and other similar theories.
We work this out explicitly in Appendix~\ref{app:hn_theory}
(for the Hellings-Nordtvedt vector-tensor theory~\cite{Hellings:1973zz,Will:1993ns} studied in~\cite{Annulli:2019fzq})
and in Appendix~\ref{app:vgb_theory} (for the vector-Gauss-Bonnet theory
of~\cite{Ramazanoglu:2019gbz,Barton:2021wfj}).

\section{Conclusions}
\label{sec:conclusions}

We revisited the tensor-vector gravity model proposed in
Ref.~\cite{Ramazanoglu:2017xbl} and explored the
vectorization process using perturbation theory.
This was done by working with a gauge invariant, Stueckelberg version of the theory
and complemented with an analysis of the Lorenz constraint in the Proca gauge.
In analogy with scalarization, one would expect to see the vector field develops a tachyonic instability, which is then quenched nonlinearly, and this process gives rise to the vectorized configurations found in previous work.
Instead, we have uncovered a ghost instability. This results demonstrates quite clearly that the strong resemblance of this model of vectorization to the Damour--Esposito-Far\`ese model of scalarization is in fact rather misleading and a phase transition process that is physically similar to scalarization does not take place.

A potential way out may exist if one can tame the ghost instability nonlinearly,
similar to the quenching of the tachyonic instability in scalarization.
Indeed, ``ghost-based spontaneous tensorization'' has been investigated~\cite{Ramazanoglu:2017yun}.
In the vectorization model studied here, ghosts appear inadvertently, and
there is no explicit derivative coupling before the introduction of the Stueckelberg mechanism. Yet, if a nonlinear quenching mechanism
exists, it could, in principle, suppress the ghost. Note that the $\hat{z}$ term in Eq.~\eqref{eq:constraint2} that controls
the instability approaches its GR value of $\hat{z}=1$ when
$A_\mu A^\mu \to \infty$ (for $\beta < 0$).
Hence, a solution with large vector field
values can lead to a case where $\hat{z}>1$ everywhere. This possibility was recently investigated for action~\eqref{eq:action} in Ref.~\cite{Demirboga2021}, and all computed static and spherically symmetric vectorized neutron stars were shown to still carry ghost or gradient instabilities. Hence, there is no sign of a quenching of the instabilities so far.

The \emph{main issue} however with the ghost instabilities we investigated is that it is not
known whether their time evolution can be done.
Even if a vector field growing to large values might quench the ghost,
it is not clear if the very time evolution of the vector field that leads to growth
can be formulated as a well-posed initial value problem due to the divergent
terms such as those in Eq.~\eqref{eq:vec_component}. The resolution of this
issue requires a mathematical analysis of the partial differential
equations we have, which is beyond the scope of this work.
We remark that these are not problems in the Proca limit of our model and
in the absence of matter, in which numerical relativity simulations have been performed,
e.g., in Refs.~\cite{Zilhao:2015tya,East:2017ovw,East:2017mrj}.

Spontaneous vectorization theories with restored gauge symmetry were
also conceived using the Higgs mechanism rather than the Stueckelberg mechanism~\cite{Ramazanoglu:2018tig}, inspired by the gravitational
Higgs mechanism~\cite{Coates:2016ktu,Franchini:2017zzx,Krall:2020kto}.
However, this theory~\cite{Ramazanoglu:2018tig} also has divergent terms in its field equations akin
to Eq.~\eqref{eq:vector_eom_mani},
hence, it is susceptible to the same
ill-posedness problems we discussed here.

We worked on the specific theory of Eq.~\eqref{eq:eom_vt}, but
other spontaneous vectorization models in the literature have similar field
equations where $\nabla_\mu F^{\mu \alpha}$ directly appears as the principal
part~\cite{BeltranJimenez:2013fca,Annulli:2019fzq,Ramazanoglu:2019gbz,Ramazanoglu:2019jrr,Kase:2020yhw,Barton:2021wfj}.
Hence, a constraint can be obtained the same way as we did, which leads to divergent terms using the arguments in Sec.~\ref{sec:nonlinear} or related ones, as we show in Appendices~\ref{app:hn_theory} and~\ref{app:vgb_theory}.

Lastly, we stress that our results are relevant for most known
extensions of spontaneous scalarization to other fields, not just the
vectors, and our study can be considered as a first step to obtain
a no-go theorem for extending spontaneous
scalarization to other fields.
For vector fields, Garcia-Saenz et al.~\cite{Garcia-Saenz:2021uyv}
has identified the presence of ghost and gradient instabilities in
the background of compact objects in a broad class of generalized Proca theories~\cite{Tasinato:2014eka,Heisenberg:2014rta}.
Similar concerns were also raised in the context of cosmology in Ref.~\cite{Esposito-Farese:2009wbc}.
Going beyond vector fields,
all known formulations of nonminimally coupled spin-2 fields
that could spontaneously grow are known to lead to ghost instabilities
as well~\cite{Ramazanoglu:2017yun}.
Likewise, $p$-form fields also have the same
constraint structure we discussed in
Sec.~\ref{sec:nonlinear}, hence
they suffer from similar divergent terms~\cite{Ramazanoglu:2019jfy}.
Spontaneous growth of spinor fields as it was introduced in Ref.~\cite{Ramazanoglu:2018hwk}
also contains divergent terms.

The only potential exception to our long list of problematic theories is a second form of spontaneous spinorization theory proposed in Ref.~\cite{Minamitsuji:2020hpl}, whose equations of motion are not known to feature divergences.
It remains to be seen if other well-posed theories exist. If this is the case, understanding what distinguishes these theories at a fundamental level from the problematic ones may lead to a proper no-go theorem for arbitrary generalizations of spontaneous scalarization.

\acknowledgments

We thank Leonardo Gualtieri, Kirill Krasnov, Helvi Witek and
Jun Zhang for discussions.
T.P.S. acknowledges partial support from the STFC Consolidated Grants
No.~ST/T000732/1 and No.~ST/V005596/1.
F.M.R. and A.C. were supported by Grant No.~117F295 of the Scientific and
Technological Research Council of Turkey (T\"UB\.ITAK).
A.C. acknowledges financial support from the European Commision and T\"UB\.ITAK
under the CO-FUNDED Brain Circulation Scheme 2, Project No.~120C081.
F.M.R. is also supported by a Young Scientist (BAGEP) Award of Bilim Akademisi
of Turkey.
H.O.S. thanks the hospitality of the University of Nottingham where
this work started.
The authors also acknowledge networking support by the GWverse COST Action
CA16104, ``Black holes, gravitational waves and fundamental physics''.
Some of our calculations were performed with the {\sc Mathematica}
packages {\sc xPert}~\cite{Brizuela:2008ra} and {\sc xCoba},
parts of the {\sc xAct/xTensor} suite~\cite{Mart_n_Garc_a_2008,xAct}.

\appendix

\section{The Hellings-Nordtvedt theory}
\label{app:hn_theory}

In this Appendix we apply the approach of Sec.~\ref{sec:nonlinear}
to examine the field equations in the
Hellings-Nordtvedt~\cite{Hellings:1973zz,Will:1993ns} vector-tensor theory
studied in Ref.~\cite{Annulli:2019fzq} as a vectorization model.

In this theory, the vector field obeys the field equation,
\be
\nabla_{\mu} F^{\mu \ag}
- \frac{1}{2} \omega R A^{\ag}
- \frac{1}{2} \eta R^{\ag}_{\spc\mu} A^{\mu} = 0,
\label{eq:hn}
\ee
where $\omega$ and $\eta$ are dimensionless coupling constants.

Let us first obtain a generalized Lorenz constraint satisfied by $A^{\ag}$ by taking a covariant derivative of Eq.~\eqref{eq:hn} and using $\nabla_{\nu} \nabla_{\mu} F^{\nu \mu}= 0$,
\be
\nabla_{\mu} (\omega R A^{\mu} + \eta R^{\mu}_{\spc\nu} A^{\nu}) = 0.
\label{eq:hn_lorentz_no_einstein}
\ee
We can expand this equation and replace the Ricci tensor with
the Einstein tensor and the Ricci scalar. The resulting constraint equation is,
\begin{align}
\nabla_{\mu}A^{\mu}
+ A^{\mu} \nabla_{\mu} \ln |\omega R|
+ \frac{2\eta}{\eta + 2\omega} \frac{1}{R} G^{\mu}_{\,\,\nu} \nabla_{\mu} A^{\nu} = 0.\nn \\
\label{eq:hn_lorentz}
\end{align}

We can now return to Eq.~\eqref{eq:hn}, write $F^{\ag\bg}$ in terms
of $A^{\ag}$, follow the same steps that lead to Eq.~\eqref{eq:vector_eom_mani}, and
find:
\be
\Box A_{\ag}  - \nabla_{\ag} \nabla_{\mu} A^{\mu} - \left[\frac{1}{2} \omega R g_{\ag \mu} + \left(1 + \frac{\eta}{2}\right) R_{\ag\mu}\right] A^{\mu} = 0.
\ee
At last, using Eq.~\eqref{eq:hn_lorentz} we obtain,
\begin{align}
\Box A_{\ag} &+ \nabla_{\mu} \ln(|\omega R|) \nabla_{\ag}A^{\mu}
+ \nabla_{\ag}\left( \frac{2\eta}{\eta + 2\omega} \frac{1}{R} G^{\mu}_{\,\,\nu} \nabla_{\mu} A^{\nu}\right)
\nn \\
&-\, {\cal M}_{\ag \mu} A^{\mu} = 0,
\label{eq:hn_wavelike_equation}
\end{align}
where
\be
{\cal M}_{\ag \bg} = \frac{1}{2} \omega R g_{\ag \bg} + \left(1 + \frac{\eta}{2}\right) R_{\ag\bg} - \nabla_{\ag} \nabla_{\bg} \ln |\omega R|,
\ee
which should be compared against Eq.~\eqref{eq:mass-tensor}.
Note that in Eq.~\eqref{eq:hn_wavelike_equation} the last term in the first line is also second order, hence, it contributes to the principal part of the differential equation in addition to the wave operator. Hence, this equations is not in an explicitly hyperbolic form, and we cannot immediately identify ${\cal M}_{\ag \bg}$ as a squared-mass tensor whose eigenvalues are related to the effective masses of the individual degrees of freedom.
However, such identification is possible in the special case $\eta = 0$
in which the problematic term vanishes and then:
\begin{align}
    {\cal M}_{\ag\bg}^{(\eta = 0)} &= (\omega / 2) R g_{\ag \bg} + R_{\ag\bg} - \nabla_{\ag} \nabla_{\bg} \ln |\omega R|.
    \label{eq:mass-tensor-hn-eta0}
\end{align}
We see, by comparing with Eqs.~\eqref{eq:mass-tensor-hn-eta0} and~\eqref{eq:mass-tensor}, that $\omega R$ plays the role of $\hat{z}$.
We then conclude that a ghost arises for the same
reasons discussed in Sec.~\ref{sec:nonlinear}.

For the general case $\eta \neq 0$ it is more convenient to analyse
the constraint~\eqref{eq:hn_lorentz_no_einstein} which we write as,
\be
\label{eq:hn_const_general}
\nabla_{\alpha} \left( \Xi^{\alpha}{}_{\beta} A^{\beta} \right) = 0,
\ee
where
\begin{equation}
\Xi^{\ag}{}_{\bg} = \eta \, G^{\ag}{}_{\bg}
+ \left(\omega + \eta/2 \right) R \, \delta^{\ag}{}_{\bg}.
\label{eq:Xi}
\end{equation}

Let us focus on the perturbative regime where the background metric is fixed and the Einstein equations hold, i.e., $G_{\ag\bg} = 8\pi T_{\ag\bg}$~\cite{Annulli:2019fzq}.
For a static, spherically symmetric perfect fluid star with energy density $\varepsilon$ and pressure $p$,
\begin{align}
\Xi^\alpha {}_{\beta} &=
4\pi \eta [
(\varepsilon-p)\delta^{\ag}{}_{\bg}
-2(\varepsilon+p) \delta^{\ag}{}_{0}\, \delta^{0}{}_{\bg}]
\nonumber\\
&\quad +8\pi \omega (\varepsilon-3p) \delta^{\ag}{}_{\bg},
\end{align}
which is diagonal.
The constraint can then be written as
\be
\partial_{0} (\sqrt{-g}\ \Xi^{0}{}_{0} A^0) = -\sum_{k} \partial_{k} (\sqrt{-g}\ \Xi^{k}{}_{k} A^{k}),
\ee
where we wrote the summation over the spatial coordinates $k$ explicitly to avoid confusion.
This means the diagonal elements have the role of a generalized $\hat{z}$ in the massless case in Eq.~\eqref{eq:def_zhat2}. We see that $\partial_0 A^0$ has a contribution in the form of
\be
\partial_0 A^0 = -\frac{\partial_r
(\Xi^{r}{}_{r})}{\Xi^{0}{}_{0}} A^r +\dots
\ee
The behavior of this term is given by the dependence of the energy density and the pressure on the radial coordinate at the surface of the star. We normally encounter power law dependence in stars due to the TOV equations as we mentioned in relation to Eq.~\eqref{eq:def_zhat2}.
Hence, $\partial_0 A^0$ diverges for generic configurations of $A^\mu$.

We conclude by noticing that the constraint equations of
disformally coupled vector-tensor theories of Ref.~\cite{Ramazanoglu:2019jrr,Minamitsuji:2020pak}  have a similar structure to
Eq.~\eqref{eq:hn_const_general}, which would lead to similar results in terms
of divergences.

\section{Vector-Gauss-Bonnet theory}
\label{app:vgb_theory}

In this Appendix we apply the approach of Sec.~\ref{sec:nonlinear} to examine
the field equations in the vector-Gauss-Bonnet theory
introduced in Ref.~\cite{Ramazanoglu:2019gbz}, and further studied in Ref.~\cite{Barton:2021wfj}.
The motivation behind these theories is to generalize the spontaneous scalarization of black holes~\cite{Doneva:2017bvd,Silva:2017uqg} to vector fields.

In this theory, the vector field obeys the field equation,
\be
\nabla_{\mu}F^{\mu\ag} = v A^{\ag} - f\, {\mathscr G} A^{\ag},
\ee
with
\be
v = \frac{1}{2} \frac{\dd V(A_{\mu}A^{\mu})}{\dd (A_{\mu}A^{\mu})},
\quad
f = \frac{1}{2} \frac{\dd F(A_{\mu}A^{\mu})}{\dd (A_{\mu}A^{\mu})},
\ee
where $V$ is the vector field's self-interaction potential, $F$ prescribes
the coupling between the vector field and ${\mathscr G}$, the Gauss-Bonnet invariant.
For small field perturbations, the potential $V$ and coupling function $F$ considered
in Ref.~\cite{Barton:2021wfj} reduce to:
\be
v = \mv^2,
\quad
f = \beta / 2,
\ee
where $\mv^2$ is the bare mass of $A^{\ag}$ and $\beta$ a coupling constant.
As with Eq.~\eqref{eq:eom_vt}, one can identify an ``effective mass squared''
$\meff^2 = \hat{z} \mv^2$, but where now $\hat{z} = 1 - (\beta / \mv^2) \, {\mathscr G} / 2$.

We can now proceed in the same manner as in Sec.~\ref{sec:nonlinear} to obtain
\be
\Box A_{\ag} + (\nabla_{\mu}\ln |\hat{z}|) \nabla_{\ag} A^{\mu} - {\cal M}_{\alpha\mu}A^{\mu} = 0,
\ee
where
\be
{\cal M}_{\ag\bg} = \mv^2 \hat{z} g_{\ag\bg} - \nabla_{\bg} \nabla_{\ag}\ln|\hat{z}|,
\ee
[compare against Eqs.~\eqref{eq:vec_component}--\eqref{eq:mass-tensor}]
where the absence of the Ricci tensor is due to the
assumption of the GR background being Ricci flat~\cite{Barton:2021wfj}.
Therefore, this theory suffers from a ghost instability as the one considered
in Ref.~\cite{Ramazanoglu:2017xbl}.

For a Schwarzschild black hole, ${\mathscr G}$ is positive everywhere. Thus, for $v=0$ the above argument cannot be repeated verbatim.
However, ${\mathscr G}$, and hence the effective mass squared, changes sign in some regions outside the event horizon of black holes with dimensionless spin $\gtrsim 0.5$~\cite{Cherubini:2002gen}.
Therefore, these commonly encountered astrophysical systems lead to divergent field equations
in such theories.

Neutron stars also feature divergences for the case of $v=0$. On a fixed general relativistic background, the Gauss-Bonnet invariant of a static spherically symmetric perfect fluid star of energy density $\varepsilon$ and pressure $p$ is given by~\cite{Silva:2017uqg}
\be
{\mathscr G}(r) = \frac{48 \mu^2}{r^6}
-128 \pi \left(2\pi p + \frac{\mu}{r^3} \right) \varepsilon,
\ee
which is positive definite outside the star.
On the other hand, near the center of the star $r = r_{c}$, the TOV equations imply
that the mass function is approximately
\be
\mu_{c} = \mu(r_c) \approx \tfrac{4}{3} \pi \varepsilon_{c} r_{c}^3,
\ee
where $\varepsilon_c = \varepsilon(r_c)$ is the central energy density.
We then find that in the star's center,
\be
{\mathscr G}_{c} \approx -256\pi^2 \left(p_c + {\varepsilon_c}/{3} \right)\varepsilon_c,
\label{eq:G_c}
\ee
where $p_c = p(r_c)$ is the central pressure.
The right hand side of Eq.~\eqref{eq:G_c} is negative
meaning that ${\mathscr G}$, and thus $\mu^{2}_{\rm eff}$, change sign within the star,
numerically found to happen near the surface~\cite{Silva:2017uqg}.

\bibliography{biblio}

\end{document}